\documentclass[aps, superscriptaddress, a4paper, twocolumn, nofootinbib, showpacs, amsfonts, amsmath,amssymb]{revtex4}

\usepackage{graphicx}
\usepackage{dcolumn}
\usepackage{bm}

\usepackage{hyperref}

\def\be{\begin{equation}}
\def\ee{\end{equation}}
\def\bea{\begin{eqnarray}}
\def\eea{\end{eqnarray}}

\begin{document}

\preprint{APS/123-QED}

\title{A novel phenomenological parametrization of dynamical dark energy via matter-gravity coupling}

\author{Hrishikesh Chakrabarty}
\email{hchakrabarty@nuaa.edu.cn}
\affiliation{Center for the Cross-disciplinary Research of Space Science and Quantum-technologies (CROSS-Q), College of Physics, Nanjing University of Aeronautics and Astronautics, 29 Jiangjun Road, Nanjing City, Jiangsu Province 211106, China}
\affiliation{Department of Physics, School of Sciences and Humanities, Nazarbayev University, Kabanbay Batyr 53, 010000 Astana, Kazakhstan}

\author{Daniele Malafarina}
\email{daniele.malafarina@nu.edu.kz}
\affiliation{Department of Physics, School of Sciences and Humanities, Nazarbayev University, Kabanbay Batyr 53, 010000 Astana, Kazakhstan}

\date{\today}

\begin{abstract}
We develop a new parametrization for generic departures from the $\Lambda$CDM model based on the Markov-Mukhanov action with non-minimal matter-gravity coupling. In this approach the cosmological constant $\Lambda$ and the gravitational coupling $G$ become dynamical with their values determined by the explicit form of the matter-gravity coupling. We use the framework to build a simple phenomenological model where dynamical dark energy is naturally obtained as a consequence of the departure of the theory from classical General Relativity at low densities. We show how the model's free parameters can be constrained from observational data and interpret the tentative observations of `phantom crossing' by the DESI collaboration in view of the proposed formalism.

\end{abstract}

\maketitle

\section{Introduction}\label{sec1}

The standard model of cosmology, known as the $\Lambda$CDM model has been very successful in describing the observable universe with very few model parameters and relying solely of its geometrical structure and matter content. The main components of the current Universe, as described by the $\Lambda$CDM model, are cold dark matter (CDM) and the cosmological constant $\Lambda$ which is a form of dark energy  \cite{Padmanabhan:2002ji,Peebles:2002gy}, along with negligible amounts of radiation and curvature. The model has stood the test of time, confirmed by precision observation of the cosmic microwave background (CMB) \cite{Planck:2018vyg}, type Ia supernovae \cite{Brout:2022vxf,Riess:2021jrx,2012ApJ...746...85S}, galaxy clustering and large scale structures \cite{DESI:2024mwx,BOSS:2016wmc,eBOSS:2020yzd,Addison:2013haa}. The remarkable feature of the $\Lambda$CDM model is that it is internally consistent across multiple probes.

Despite its successes, the models still suffers from a number issues both from the theoretical and observational sides. The major component that dominates the current universe's expansion is the cosmological constant $\Lambda$, which was introduced by Einstein in the early 20th century in order to construct his static model of the universe \cite{Einstein:1917ce}. 
While Einstein's $\Lambda$ became unnecessary after Hubble's discovery of the universe's expansion, a new form of cosmological constant, named `dark energy' had to be introduced in 1998 following 
the observation that the universe is currently going through a phase of accelerated expansion \cite{SupernovaSearchTeam:1998fmf}. 
However, at present, the nature of dark energy and a solid theoretical framework to explain its existence are still missing. 
A discussion on various methods to solve the cosmological constant problem can be found in \cite{Carroll:2000fy}.

At the same time, recent observational developments, such as measurements from the Dark Energy Spectroscopic Instrument (DESI), show that dark energy 
may be dynamical in nature \cite{DESI:2024mwx,DESI:2025fii}. The standard $\Lambda$CDM model predicts a constant equation of state, meaning the dark energy density remains constant over time. On the other hand, DESI's DR I and DR II results hint at a statistically significant redshift dependence of the equation of state. When the data is confronted with the simplest parametrized dark energy model, known as $w_0w_a$CDM \cite{Chevallier:2000qy,Linder:2002et}, the deviation from a constant equation of state becomes significant as the results exclude the $\Lambda$CDM model at $\sim 2\sigma - 4\sigma $ depending on different dataset combinations \cite{DESI:2024mwx}. This exclusion of $\Lambda$CDM is driven by a non-zero preference for the parameter $w_a$ which tracks the dynamical nature of the equation of state close to today. Even more interestingly, the best fit values of $w_0$ and $w_a$ seem to hint towards a phantom nature of dark energy, with an equation of state parameter $w<-1$ in the past. 
This kind of behavior may be obtained from models of dark energy other than $\Lambda$, such as the ones emerging from additional scalar fields, and they can explain the dynamical nature when confronted with DESI data (see e.g. \cite{Cline:2025sbt}).  
This has led to increased interest in dynamical dark energy models (see e.g \cite{Cline:2025sbt,Capozziello:2025qmh,Carloni:2025dqt}) in recent years. 

Several phenomenological and theoretically motivated frameworks have been proposed in which the dark-energy contribution is allowed to evolve with time rather than being identified with the cosmological constant. Early examples include scalar field models, phantom scenarios, higher order curvature and modified gravity models, see e.g. \cite{Carroll:1998zi,Chiba:1999ka,Carroll:2004hc,Gannouji:2006jm,Capozziello:2005tf,Deffayet:2000uy,Deffayet:2001pu,Ludwick:2017tox}. More recently, running-vacuum or running-Einstein-dark-energy models have considered a scale- or time-dependent vacuum contribution, providing another mechanism through which the effective dark-energy density can depart from a constant value \cite{Montani:2024ejp}. These approaches belong to the broader class of dark energy models extensively reviewed earlier in \cite{Copeland:2006wr}.

In this article we adopt the framework for a simple modification of Einstein's gravity, proposed by Markov and Mukhanov in \cite{Markov:1985py} and cast it as a dynamical dark energy cosmological model. 
Markov-Mukhanov's theory relies on a
non-minimal matter-gravity coupling that depends only on the energy scale and naturally leads to dynamical Newton's and cosmological constants. The behavior of the variable $G$ and $\Lambda$ then depends on the nature of the coupling. 
We introduce late-time corrections to the $\Lambda$CDM 
by assuming an expansion at low densities of the matter-gravity coupling close to today and test the resulting induced dynamical dark energy component via combinations of observational data from DESI \cite{DESI:2024mwx}, Planck \cite{Planck:2018vyg,Lemos:2023xhs}, cosmic chronometers \cite{Jimenez:2001gg,Borghi:2021rft} and supernovae surveys \cite{Brout:2022vxf,Scolnic:2021amr}.

The article is organized as follows: In Sec.~\ref{sec2}, we discuss the Markov-Mukhanov model starting from the action and derive the relevant equations for cosmology. In Sec.~\ref{sec3}, we introduce the late-time corrections to the $\Lambda$CDM model within Markov-Mukhanov's framework and discuss the possible choices on the expansion terms and their properties, thus proposing two phenomenological models. 
In Sec.~\ref{sec4}, we test the models with observational data and put constrains on the models parameters. Finally in Sec.\ref{sec5}, we summarize and discuss our results. Throughout the article we use the metric signature $(-,+,+,+)$ and natural units $c = \hbar = 1$, unless otherwise stated.

\section{Cosmology from the Markov-Mukhanov action}\label{sec2}

In this section we shall introduce the main formalism for the Markov-Mukhanov (MM) theory.  
We start with the action \cite{Markov:1985py} 
\begin{equation}\label{eq-action}
    S = \int d^4x \sqrt{-g} \left( \frac{R}{8\pi G_N} + 2\chi (\varepsilon) \mathcal{L}_{\rm m} \right),
\end{equation}
where $R$ is the Ricci scalar and $ \mathcal{L}_{\rm m} $ is the matter Lagrangian. 
We can choose the matter Lagrangian at will in order to model the gravitational sources in the theory and in this article we follow the standard prescription for cosmology adopting a perfect fluid description for normal matter with $\mathcal{L}_m=\varepsilon$, where $\varepsilon$ is the fluid's energy density.
Notice that in the above action we have a non-minimal matter-gravity coupling that depends on the energy scale as described by the free function $\chi(\varepsilon)$. Also notice that the exact form of the coupling can be obtained from a more fundamental theory of gravity such as, for example, Asymptotic Safety as done in \cite{Zholdasbek:2024pxi} or via phenomenological considerations as done in \cite{Chakrabarty:2025gha,Zhumabek:2024tvp}.

The action \eqref{eq-action} can be understood as an effective classical description of a theory describing departures from General Relativity (GR) at higher and lower densities with respect to the regime where GR works. In terms of cosmology this is equivalent to departures from GR at early or late-times. We can then interpret such departures as originating from a more fundamental theory of gravity or quantum gravity. On the other hand, we may also view the MM formalism as a model agnostic way to describe classical departures from GR. In this manner, we can aim at determining the properties of such a theory phenomenologically by finding the function $\chi(\varepsilon)$ that is preferred by observations. It is clear from equation \eqref{eq-action} that for $\chi=1$ one obtains classical GR without the cosmological constant. At the same time it is evident that $\Lambda$ appears when choosing $\chi=1+\varepsilon_\Lambda/\varepsilon$, with $\varepsilon_\Lambda$ constant, which suggests that the cosmological constant may be merely the lowest order approximation of $\chi$ at low densities. This is the approach we aim to take in the following.

The variation of the action with respect to the metric tensor leads to the modified Einstein equation as
\begin{equation}\label{eq-ee}
    R_{\mu\nu} - \frac{1}{2}g_{\mu\nu}R = 
   8\pi G_N \tilde{T}_{\mu\nu},
\end{equation}
where
\begin{equation}\label{eq-eff-emt}
    \tilde{T}_{\mu\nu} = \left( \varepsilon\chi \right)_{,\varepsilon}T_{\mu\nu} + (\varepsilon^2 \chi_{,\varepsilon}) g_{\mu\nu},
\end{equation}
is the effective energy-momentum tensor and 
\begin{equation}\label{Tmunu}
    T_{\mu\nu} = \left( \varepsilon + P(\varepsilon) \right)u_\mu u_\nu + P(\varepsilon) g_{\mu\nu},
\end{equation}
is the usual energy-momentum tensor for a perfect fluid with $P (\varepsilon)$ the fluid's pressure. The pressure is related to the energy density $\varepsilon$ via a linear equation of state $P=w\varepsilon$, and $u^\mu$ is the fluid's four velocity.
As we can see from Eq.~\eqref{eq-ee}-\eqref{eq-eff-emt}, the MM model can be interpreted as a model with variable Newton's and cosmological constants. In fact, looking at the first and second terms in \eqref{eq-eff-emt}, we can define 
\begin{equation}
G(\varepsilon) = G_N(\chi \varepsilon)_{,\varepsilon} \; \; \text{and} \; \; \Lambda(\varepsilon) = - 8\pi G_N \varepsilon^2 \chi_{,\varepsilon} 
\end{equation}
as the running Newton's constant $G(\varepsilon)$ and the running cosmological constant $\Lambda(\varepsilon)$. 
However, it is important to remark that the Einstein equation can also be written as GR plus a correction term where
\be \label{T-2fluids}
\tilde{T}_{\mu\nu} = T_{\mu\nu}+T^{\rm corr}_{\mu\nu},
\ee 
and the effects of the matter-gravity coupling $\chi$ are all absorbed into the correction term
\be 
T^{\rm corr}_{\mu\nu} = \Big[\left( \varepsilon\chi \right)_{,\varepsilon}-1\Big]T_{\mu\nu} + (\varepsilon^2 \chi_{,\varepsilon}) g_{\mu\nu}.
\ee 
Then $T^{\rm corr}_{\mu\nu}$ can be interpreted as an additional fluid component in GR and its presence may be used to model the observed dynamical dark energy.

To ensure consistency of the field equations with the effective matter source $\Tilde{T}^\mu_\nu$, the effective stress-energy tensor must satisfy the standard conservation law, i.e.
\begin{equation}\label{eq-bianchi}
    \nabla_\mu \Tilde{T}^\mu_\nu = 0.
\end{equation}
We can project the above equation along the four-velocity field $u^\mu$ to see how the effective energy density evolves, this gives 
\begin{equation}\label{tilde-cons}
    \nabla_\mu \left(\Tilde{\varepsilon}u^\mu \right) + \Tilde{P}\nabla_\mu u^\mu = 0,
\end{equation}
where $\tilde{\varepsilon}$ and $\tilde{P}$ are the effective fluid's energy-density and pressure that can be obtained from Eq.~\eqref{eq-eff-emt} as 
\begin{equation} \label{epsilon-tilde}
    \begin{aligned}
        \tilde{\varepsilon} &=  (\chi \varepsilon)_{,\varepsilon}\varepsilon  -\varepsilon^2 \chi_{,\varepsilon} = \varepsilon\chi(\varepsilon), \\
        \tilde{P} &= (\chi \varepsilon)_{,\varepsilon}P+\varepsilon^2 \chi_{,\varepsilon}.
    \end{aligned}
\end{equation}
Now, using these expressions in Eq.~\eqref{tilde-cons} we obtain
\begin{align}
    \left(\varepsilon\chi\right)_{,\varepsilon}\left[\nabla_\mu \left(\varepsilon u^\mu \right) + P\nabla_\mu u^\mu\right] = 0,
\end{align}
which is equivalent to the continuity equation for the classical fluid, provided that $\left(\varepsilon\chi\right)_{,\varepsilon}\neq 0$. Therefore, the conservation of the classical energy density $\varepsilon$ follows from the conservation of effective energy density $\tilde{\varepsilon}$ (since for us $\left(\varepsilon\chi\right)_{,\epsilon} \neq 0$) \cite{Zholdasbek:2024pxi}.

We can now define an effective equation of state assuming the relation $\tilde{P}=\tilde{w}\tilde{\varepsilon}$ where the variable equation of state parameter $\tilde{w}(\varepsilon)$ is given by
\be \label{tilde-w}
\tilde{w}=-1+\frac{(\chi \varepsilon)_{,\varepsilon}}{\chi}\left(1+\frac{P}{\varepsilon}\right) = w+(1+w)\frac{\varepsilon \chi_{,\varepsilon}}{\chi}.
\ee
Notice that we retrieve GR without cosmological constant by choosing $\chi=1$ as $G(\varepsilon)=G_N$, $\Lambda(\varepsilon)=0$ and $\tilde{w}=w=P/\varepsilon$. Similarly it is easy to see that GR with the cosmological constant may be obtained by choosing $\chi=1+\varepsilon_\Lambda/\varepsilon$ as we get $G(\varepsilon)=G_N$ and $\Lambda(\varepsilon)=8\pi G_N \varepsilon_\Lambda$.

Finally, from Eq.~\eqref{T-2fluids} we may treat the model as two interacting fluids in GR with one of them $T^{\mu\nu}=T_m^{\mu\nu}$ given by dust (for which $\varepsilon_m=\varepsilon$ and $P_m=0$) and the other $T_{\rm corr}^{\mu\nu}=T_{de}^{\mu\nu}$ describing the dynamical dark energy component with density $\varepsilon_{de}$. We then have
\begin{equation} \label{eq-2fluids}
    \begin{aligned}
        \tilde{\varepsilon} &= \varepsilon_m + \varepsilon_{de} =  \varepsilon + \varepsilon(\chi(\varepsilon)-1), \\
        \tilde{P} &= P_m + P_{de} =0+\varepsilon^2 \chi_{,\varepsilon}.
    \end{aligned}
\end{equation}

\subsection{Cosmology}

Let us now apply the MM formalism to cosmology. The evolution of the universe is governed by the Friedmann equations which in our case will be derived from the MM modification of Einstein's equation. We start by considering a FRW universe with the metric given by
\begin{equation}\label{eq-metric}
    ds^2 = -dt^2 + a^2(t)\left(\frac{dr^2}{1-kr^2} + r^2d\Omega^2\right),
\end{equation}
where $a(t)$ is the scale factor, $k$ is the curvature and $d\Omega^2$ is the line element on the unit 2-sphere. We can then derive the Friedmann equations for the general MM action with the effective energy-momentum tensor \eqref{eq-eff-emt} as
\begin{equation} \label{friedmann}
    \begin{aligned}
        H^2 &= \left(\frac{\dot{a}}{a}\right)^2 = \frac{8\pi G_N}{3}\varepsilon \chi - 
        \frac{k}{a^2} = \frac{8\pi G_N}{3}\tilde{\varepsilon} - \frac{k}{a^2}, \\
        \frac{\ddot{a}}{a} &= -\frac{4\pi G_N}{3}\left[ (\varepsilon+3P)\chi + 3\varepsilon \chi_{,\varepsilon} \left( \varepsilon + P \right) \right] =\\
        &= -\frac{4\pi G_N}{3} (\tilde{\varepsilon}+3\tilde{P}),
    \end{aligned}    
\end{equation}
where $ H = \dot{a}/a $ is the Hubble parameter and the dot ( $\dot{}$ ) represents derivatives with respect to time. The evolution of the effective energy density can be easily obtained from the $\nu=0$ component of Eq.~\eqref{eq-bianchi} as
\begin{equation}
    \dot{\Tilde{\varepsilon}} + 3\frac{\dot{a}}{a}\left(\Tilde{\varepsilon} + \Tilde{P} \right) = 0,
\end{equation}
which simplifies in terms of the original fluid quantities as
\begin{equation}\label{eq-conservation}
    \left( \varepsilon \chi \right)_{,\varepsilon}\left[ \dot{\varepsilon} + 3\frac{\dot{a}}{a}\left(\varepsilon + P \right) \right] = 0.
\end{equation}
Since $\left( \varepsilon \chi \right)_{,\varepsilon} \neq 0$ at all times, the conservation of the classical energy density $\varepsilon$ follows from the conservation of the effective energy density $\tilde{\varepsilon}$ for an homogeneous perfect fluid, as was argued in the previous section. 

For simplicity and ease of calculation, we will assume a flat universe by setting $k=0$. We will also assume a universe with only pressureless matter (which is a valid approximation for late-time models) and by construction there will be an induced dark energy component that depends on the choice of the MM coupling $\chi$ as defined in Eq.~\eqref{eq-2fluids}. We can then cast the MM version of the first Friedmann equation as the corresponding equation for a two fluids model in GR in the following way
\begin{align}
    H^2 
    &= \frac{8\pi G_N}{3}\left( \varepsilon_m + \varepsilon_{de} \right),
\end{align}
where $\varepsilon_m$ is the energy density of the dust matter component and $\varepsilon_{de} = \varepsilon_m(\chi-1)$ represents the energy density of the induced dark energy component. Now, dividing both sides of the equation by today's value of the Hubble parameter $H_0^2$, we can obtain the dimension-less version of the Friedmann equation
\begin{align}
    \frac{H^2}{H_0^2} = \frac{\Omega_m}{a^3} + \frac{\Omega_m}{a^3}\left( \chi - 1 \right).
\end{align}
Here we have used the solution of the conservation equation above $\varepsilon_m = \Omega_m\varepsilon_0\chi_0a^{-3}$ with the scale factor today $a_0 = 1$, $\Omega_m$ is the fraction of matter energy density today and $\varepsilon_0\chi_0 = 3H_0^2/8\pi G_N$ is the total energy density of the universe today. Notice that, for $\chi=1$, we retrieve a universe with matter only as the induced dark energy component vanishes.

The equation of state for the induced dark energy component $w_{de}=P_{de}/\varepsilon_{de}$ can be written as
\begin{align}
    w_{de} &= \frac{\varepsilon\chi_{,\varepsilon}}{\chi-1}=-\frac{a}{3}\frac{\chi(a)_{,a}}{\chi(a)-1},
\end{align}
where we have used the solution of the Eq.~\eqref{eq-conservation} for dust like matter ($w=0$).

\section{Matter-Gravity induced Dynamical Dark Energy}\label{sec3}

\begin{figure*}[]
   \begin{center}
        \includegraphics[width=8.5cm]{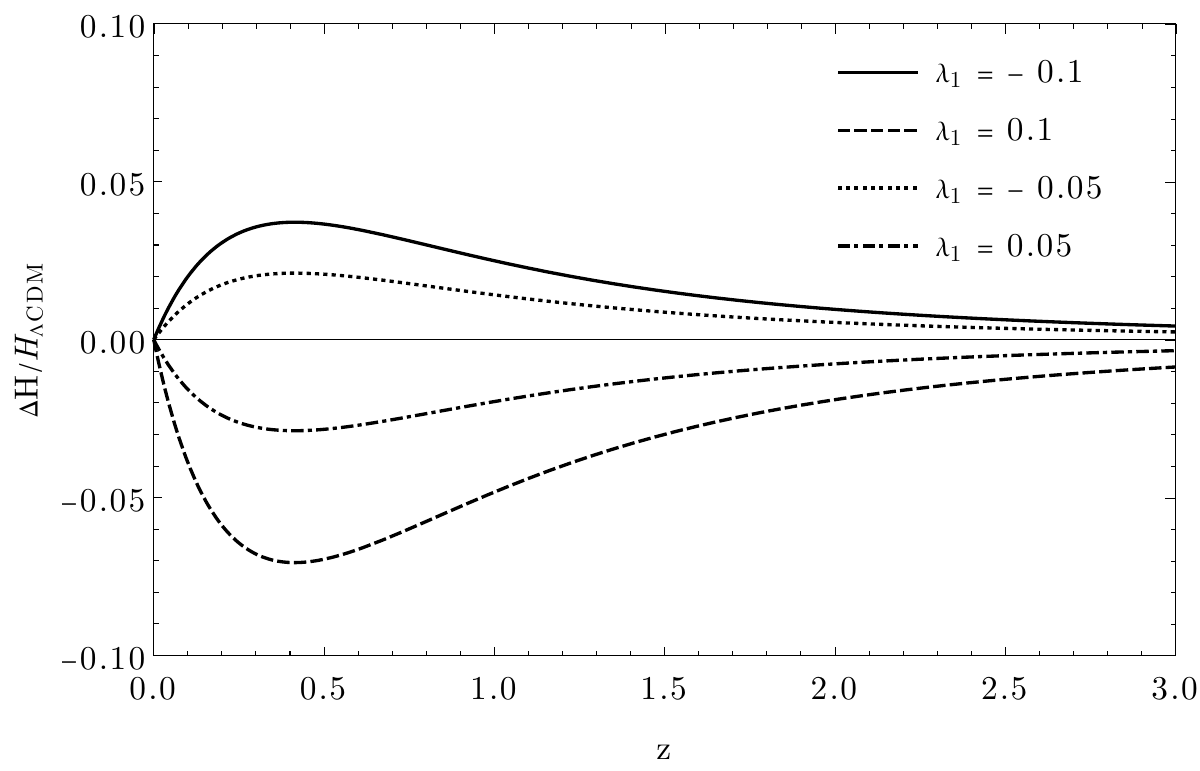}
        \includegraphics[width=8.5cm]{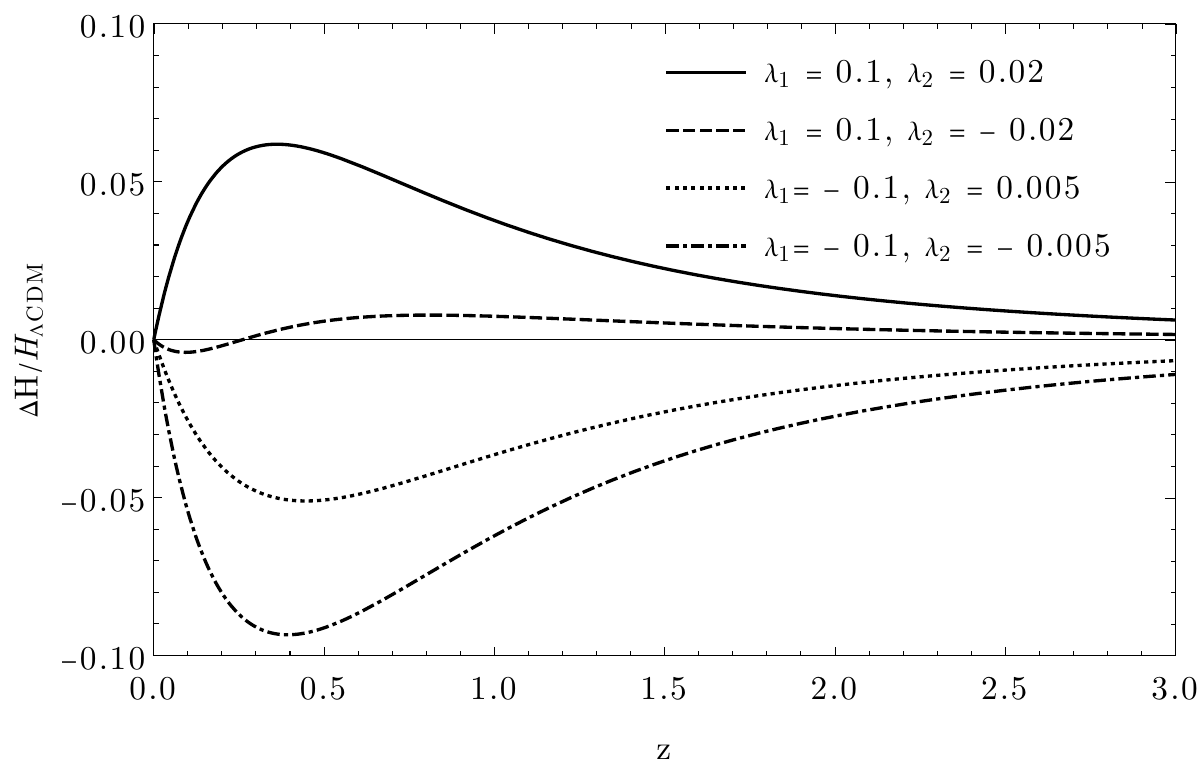}
  \end{center}
    \caption{Normalized Hubble rates, $1 - H_{\rm model}/H_{\rm \Lambda CDM}$, for $\lambda_1$CDM (left) and $\lambda_1\lambda_2$CDM (right) as a function of the redshift $z$ respectively, for different values of the model parameters. Here we have used $\Omega_m = 0.3$ for illustrative purpose.}\label{fig-Hubble}
\end{figure*}

We already mentioned how the Friedmann equations \eqref{friedmann} reduce to that of a universe with pressureless matter and cosmological constant if we choose $\chi = 1 + \varepsilon_\Lambda/\varepsilon$ where $\varepsilon_\Lambda$ is a constant. It is straightforward to check that in this case the first Friedmann equation becomes
\begin{align}
    \frac{H^2}{H_0^2} = \frac{\Omega_m}{a^3} + \Omega_\Lambda, 
\end{align}
where $\Omega_\Lambda = \varepsilon_\Lambda/(\varepsilon_0\chi_0)$ is the fraction of dark energy density. Here we are interested in how the matter-gravity coupling may induce deviations from $\Lambda$CDM at late-times when we interpret the $\chi$ that gives the $\Lambda$CDM as the first order expansion of a more general function $\chi$ at low densities. Therefore we introduce next order perturbations to the form of $\chi (\varepsilon)$  
in the following way
\begin{align}
    \chi(\varepsilon) &= 1 + \frac{\varepsilon_\Lambda}{\varepsilon}\left( 1 + \frac{\varepsilon_{\Lambda1}}{\varepsilon} + \left(\frac{\varepsilon_{\Lambda2}}{\varepsilon}\right)^2 + \dots \right) = \\
    &=1+\frac{\varepsilon_\Lambda}{\varepsilon}+\frac{\varepsilon_\Lambda\varepsilon_{\Lambda1}}{\varepsilon^2}+\frac{\varepsilon_\Lambda\varepsilon_{\Lambda2}^2}{\varepsilon^3}+ \dots \, .
\end{align}
Then the constants $\varepsilon_{\Lambda1}$ and $\varepsilon_{\Lambda2}$ in the higher order terms account for deviations from $\Lambda$CDM at their corresponding density scales. 
This may be understood as the expansion of a more fundamental effective theory as described by $\chi$ at low densities, with the two lowest order terms corresponding to GR with a cosmological constant.

The first Friedmann equation with this parametrization for $\chi$ becomes
\begin{align}
    \frac{H^2}{H_0^2}  &= \frac{\Omega_m}{a^3} + \Omega_{de}(a), 
\end{align}
with
\begin{align}
\Omega_{de}(a) = \Omega_\Lambda\left( 1 + \frac{\lambda_1}{\Omega_m/a^3}+\frac{\lambda_2}{\left(\Omega_m/a^3\right)^2} + \dots \right),
\end{align}
where $\lambda_1 = \varepsilon_{\Lambda1}/(\varepsilon_0\chi_0)$ and $\lambda_2 = (\varepsilon_{\Lambda2}/(\varepsilon_0\chi_0))^2$ characterize matter-gravity induced deviations from cosmological constant dark energy.

Going forward, we shall concentrate on two models 
\begin{itemize}
    \item[(a)] $\lambda_1$CDM: First we would like to see if there exist any matter-gravity coupling induced first order deviations from $\Lambda$CDM. Therefore, in this model we will ignore all the higher order correction terms except $\lambda_1$. The resulting Friedmann equation becomes 
    \begin{align}
        \frac{H^2}{H_0^2} = \frac{\Omega_m}{a^3} + \Omega_\Lambda\left( 1 + \frac{\lambda_1}{\Omega_m/a^3}\right),
    \end{align}
    where we enforce the normalization $H = H_0$ at $a = 1$ to obtain $\Omega_\Lambda$ as
    \begin{align}
        \Omega_\Lambda = \frac{1-\Omega_m}{1+\lambda_1/\Omega_m}.
    \end{align}
    Obviously, we recover $\Lambda$CDM if $\lambda_1 = 0$. The dark energy equation of state for this model becomes
    \begin{align}
        w_{de} = -1 - \frac{\lambda_1}{\lambda_1+\Omega_m/a^3}.
    \end{align}
    As we can see as $a\rightarrow0$, the equation of state parameter tends to $-1$. However we may have two different behaviors depending on the sign of $\lambda_1$. For $\lambda_1>0$ the dark energy model is phantom at all times with $w_{de}\rightarrow -2$ as $a$ becomes large. On the other hand for $\lambda_1<0$ we have that $w_{de}$ grows, initially behaving as quintessence, eventually becoming positive and diverging in a finite time in the future. Therefore for the $\lambda_1$CDM model a phantom dark energy close to today may be obtained only if $\lambda_1>0$ and a phantom crossing is not allowed for any value of $\lambda_1$.
    
    \item[(b)] $\lambda_1\lambda_2$CDM: As a second model we shall keep both $\lambda_1 \neq 0$ and $\lambda_2 \neq 0$ while setting the higher order terms to zero. The Friedmann equation in this case can be written as
    \begin{align}
        \frac{H^2}{H_0^2} = \frac{\Omega_m}{a^3} + \Omega_\Lambda\left( 1 + \frac{\lambda_1}{\Omega_m/a^3}+\frac{\lambda_2}{\left(\Omega_m/a^3\right)^2}\right),
    \end{align}
    and enforcing the normalization $H = H_0$ at $a = 1$ we obtain $\Omega_\Lambda$ as
    \begin{align}
        \Omega_\Lambda = \frac{1-\Omega_m}{1+\lambda_1/\Omega_m+\lambda_2/\Omega_m^2}.
    \end{align}
    We can easily see from the Friedmann equation above that the parameters $\lambda_1$ and $\lambda_2$ will tend to be degenerate and in the next section we shall show that the current available data is not precise enough to fully break this degeneracy. However this model, with two parameters, allows for a wider range of possible behaviors as can be seen from the dark energy equation of state 
    \begin{align}
        w_{de} = -1 - \frac{2\lambda_2 + \lambda_1\Omega_m/a^3}{\lambda_2 + \lambda_1\Omega_m/a^3+\left(\Omega_m/a^3\right)^2}.
    \end{align}
    If both $\lambda_1$ and $\lambda_2$ are positive the behavior of $w_{de}$ is similar to the previous case, i.e. phantom at all times with $w_{de}\rightarrow -3$ for large $a$. Similarly when both parameters are negative, the behavior resembles the previous case (never phantom and growing as $a$ increases). However, for certain values of $\lambda_1$ and $\lambda_2$ having opposite signs the model allows for a phantom epoch (in the early universe when $\lambda_1>0$ or in the late universe when $\lambda_1<0$) with phantom crossing either in the future or in the past.  
    The redshift at phantom crossing can then be obtained by solving the equation
    \begin{align}
    2\lambda_2+\lambda_1\Omega_m(1+z)^3 = 0,    
    \end{align}
    where we have used the relation $ a = 1/(1+z) $.   
\end{itemize}
In Fig.~\ref{fig-Hubble}, we plot the normalized Hubble rate ($\Delta H/H_{\rm \Lambda CDM}$ where $\Delta H = H_{\rm \Lambda CDM} - H_{\rm model}$) for these two models for different choices of $\lambda_1$ and $\lambda_2$. The left and right panels are for $\lambda_1$CDM and $\lambda_1\lambda_2$CDM respectively and illustrate the effects of the late-time deviation on the expansion rate of both models.

\section{Constraints from observations}\label{sec4}

We shall now focus on constraining the deviations from $\Lambda$CDM with recent late-time data. As discussed earlier, we will concentrate on the two models outlined, namely  
$\lambda_1$CDM with three free parameters $\{ H_0, \Omega_m, \lambda_1\}$ and $\lambda_1\lambda_2$CDM with four parameters $\{ H_0, \Omega_m, \lambda_1, \lambda_2\}$.  
In the previous section, for simplicity, we have considered the Friedmann equations for a universe with matter and a matter-gravity induced DE component. However, for numerical purposes, in the simulations we must also include radiation in addition to the usual dust matter and induced DE components.
We use the following relation for the present-day radiation density, $\Omega_r = 2.47\times10^{-5}h^{-2}(1+0.2271N_{\rm eff})$ \cite{WMAP:2008lyn}, where $h = H_0/100$ and $N_{\rm eff}$ is the effective number of neutrino species which we set to $3.044$ \cite{Lemos:2023xhs}. We perform a parameter inference procedure on our selected models using the publicly available sampler \texttt{emcee} \cite{Foreman-Mackey:2012any} to implement Markov Chain Monte Carlo (MCMC) procedure and check the convergence of our MCMC chains using Gelman $\&$ Rubin $R - 1$ parameter \cite{Gelman:1992zz}. We consider the convergence condition to be met for our chains when $R - 1 < 0.01$.

\begin{table}[]
\centering
\begin{tabular}{lr}
\hline
\textbf{Parameters} & \textbf{Prior range} \\ \hline
$H_0$               & $[40, 100]$          \\
$\Omega_m$          & $[0.0, 0.5]$         \\
$w_0$               & $[-3.0, 1.0]$        \\
$w_a$               & $[-3.0, 2.0]$        \\
$\lambda_1$         & $[-0.5, 0.5]$        \\
$\lambda_2$         & $[-0.5, 0.5]$        \\ \hline
\end{tabular}
\caption{Free parameters and their prior ranges used in our parameter inference procedure. The parameters $\lambda_1$ and $\lambda_2$ correspond to the ones used in the models outlined here. For comparison, the parameters $w_0$ and $w_1$ correspond to dark energy equation of state parametrization for the $w_0w_a$CDM model \cite{Chevallier:2000qy,Linder:2002et}.}
\label{tab:prior}
\end{table}

In our parameter inference procedure, we use informative flat priors on the parameters as tabulated in Tab. \ref{tab:prior}. We analyze the obtained samples with the \texttt{GetDist} package \cite{Lewis:2019xzd}. The datasets we use in our analysis are described below. 

\subsection{Datasets}

\begin{figure*}[]
   \begin{center}
        \includegraphics[width=8.5cm]{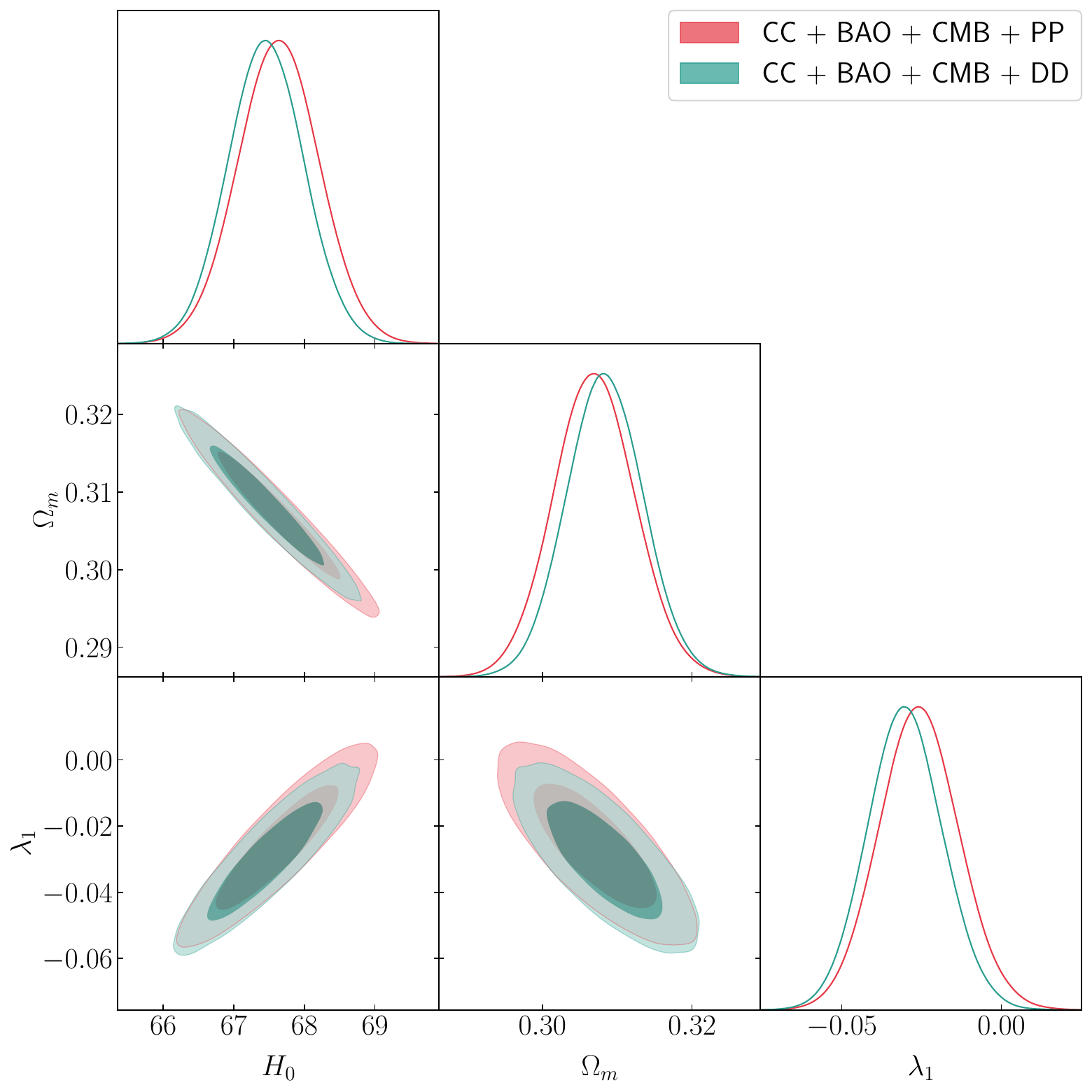}
        \includegraphics[width=8.5cm]{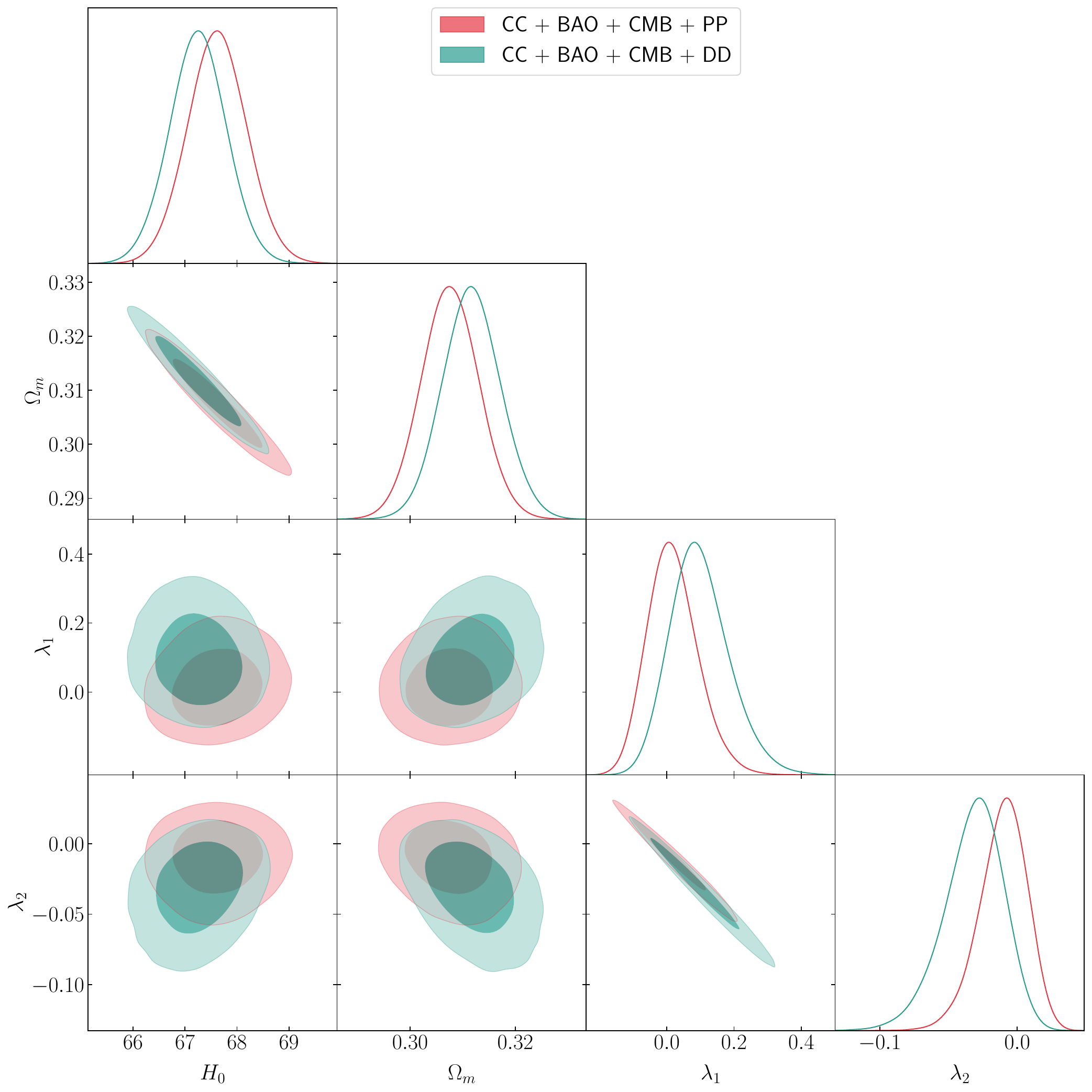}
  \end{center}
    \caption{One-dimensional posterior probability distributions and and two-dimensional 68\% and 95\% confidence level contours for the free parameters \{$H_0$, $\Omega_m$, $\lambda_1$\} of the $\lambda_1$CDM model (left panel) and \{$H_0$, $\Omega_m$, $\lambda_1$, $\lambda_2$\} of the $\lambda_1\lambda_2$CDM model (right panel), as inferred by the two dataset combinations. }\label{figcont}
\end{figure*}

\begin{table*}[]
\begin{tabular}{wl{5cm}wc{3cm}wc{3cm}wc{3cm}wc{3cm}}
\hline
Model/Dataset & $H_0$ (km/s/Mpc) & $\Omega_m$ & $w_0$ or $\lambda_1$ & $w_a$ or $\lambda_2$ \\\hline \hline 
\rule{0pt}{3ex}
\textbf{CC + BAO + CMB + DD } &  &  &  &  \\
\rule{0pt}{3ex} 
$w_0w_a$CDM & $67.59\pm0.52$ & $0.309\pm0.005$ & $-0.842\pm0.052$ & $-0.501_{-0.202}^{+0.200}$ \\
\rule{0pt}{3ex} 
$\lambda_1$CDM & $67.45\pm0.53$ & $0.308\pm0.005$ & $-0.03\pm0.011$ & $-$ \\
\rule{0pt}{3ex} 
$\lambda_1\lambda_2$CDM & $67.25\pm0.54$ & $0.311\pm0.005$ & $0.092^{+0.093}_{-0.080}$ & $-0.030^{+0.019}_{-0.022}$ \\
\\[0.1mm]
\hline
\rule{0pt}{3ex} 
\textbf{CC + BAO + CMB + PP } &  &  &  &  \\
\rule{0pt}{3ex} 
$w_0w_a$CDM & $67.71\pm0.57$ & $0.308\pm0.005$ & $-0.867\pm0.051$ & $-0.408\pm0.19$ \\
\rule{0pt}{3ex} 
$\lambda_1$CDM & $67.63\pm0.57$ & $0.306\pm0.005$ & $-0.025\pm0.012$ & $-$ \\
\rule{0pt}{3ex} 
$\lambda_1\lambda_2$CDM & $67.61\pm0.56$ & $0.307\pm0.005$ & $0.013^{+0.078}_{-0.069}$ & $-0.009^{+0.016}_{-0.018}$ \\
\\[0.1mm]
\hline
\end{tabular}
\caption{Mean values and their associated uncertainties for the inferred parameters from the MCMC parameter inference procedure for $w_0w_a$CDM, $\lambda_1$CDM and $\lambda_1\lambda_2$CDM models for two dataset combinations.}
\label{tab:med1sig}
\end{table*}

\begin{itemize}
    \item {\bf Baryon Acoustic oscillation (BAO):} We use the recent BAO measurements from DESI DR2 which includes observations of galaxies, quasars and Lyman-$\alpha$ tracers. The BAO measurements consists of the transverse comoving distance $(D_M/r_d)$, the Hubble distance $(D_H/r_d)$ and the angle averaged distance $(D_V /r_d)$ normalized to $r_d$, which is the comoving sound horizon at the drag epoch \cite{DESI:2019jxc,DESI:2024mwx,DESI:2025zgx,Moon:2023jgl}. These distances are related to the Hubble rate in the following way \cite{DESI:2025zgx} 
    \begin{align}
        \frac{D_M}{r_d} &= \frac{c}{H_0}\int_0^z\frac{dz'}{H(z)/H_0}, \\ 
        \frac{D_H}{r_d} &= \frac{c}{H(z)}, \\
        \frac{D_A}{r_d} &= \left(zD_M^2D_H\right)^{1/3},
    \end{align}
    where $r_d$ is defined as $r_d = \int_{z_d}^\infty c_s(z)/H(z)dz$. Here $c_s(z)$ is the speed of sound at the drag epoch which depends on the baryon and photon content of the Universe at that epoch. In the above equations, we restore the speed of light in vacuum $c$ and we assume a flat universe. The BAO dataset we consider has 13 data points in the redshift range $z\in[0.1, 4.16]$ and they are specifically based on the observations of the clustering of Bright Galaxy Samples (BGS), Luminous Red Galaxy Samples (LRG), Emission Line Galaxy (ELG), combined LRG and ELG, quasars and Lyman-$\alpha$ samples summarized. The dataset is summarized in Tab.IV of \cite{DESI:2025zgx}. In the MCMC procedure, we employ Hu-Sugiyama fitting formula to calculate $r_d$ \cite{Hu:1995en}. We refer to this dataset as BAO. 
    
    \item {\bf Cosmic chronometers (CC):} We use the measurement of the Hubble rate $H(z)$ by cosmic chronometers. They are  the differential ages of massive, early-time, passively evolving galaxies \cite{Jimenez:2001gg,Borghi:2021rft}. In our analysis, we adopt 15 out of more than 30 data points reported in \cite{Moresco:2012by,Moresco:2015cya,Moresco:2016mzx} that spans the redshift range $z\in[0.1791, 1.965]$. The focus is on the subset where full estimates of the covariance matrix's non-diagonal terms and systematic contributions are accessible \cite{Moresco:2018xdr,Moresco:2020fbm}. Note that, inclusion of the unused points is unlikely to affect the outcome of our analysis as the considered data points are some of the most precise and reliable measurements. We call this dataset CC.    
    \item {\bf Type Ia Supernovae:} For Type Ia Supernovae, we use two distinct datasets. First, the Pantheon+ (PP) compilation, which includes 1701 light curve measurements of 1550 uncalibrated Type Ia Supernovae in the redshift range $z\in[0.01, 2.26]$ \cite{Scolnic:2021amr}. We have ignored the SH0ES calibration in the core of this analysis, however, for completeness, we briefly mention the effect of this calibration in the results section. We call this dataset PP.

    Second, we use the DES-Dovekie sample, which presents a high-precision re-analysis of the Dark Energy Survey 5-year supernova (DES-5YSN) data. This set includes 1820 Type Ia Supernovae in the redshift range $z\in[0.02, 1.13]$ \cite{DES:2025sig}. By employing a rigorous treatment of host-galaxy dust and survey systematics than previous iterations, DES-Dovekie provides a robust geometric probe of the late-time expansion history. In our analysis, we analytically marginalize over the absolute magnitude $M$ \cite{Bridle:2001zv,Scovacricchi:2015ely,Caroli:2021mjg}. We refer to this dataset as DD.

    \item {\bf Cosmic Microwave Background (CMB):} Finally, we employ a compressed CMB likelihood in the form of correlated Gaussian priors on the quantities $\{\theta_*, \omega_b, \omega_{bc}\}$. Here $\omega_b$ and $\omega_{bc}$ are physical baryon and cold dark matter densities respectively. $\theta_* = r_*/D_M(z_*)$ is the angular scale of the acoustic fluctuations with $r_*$ and $D_M(z_*)$ being the comoving sound horizon at recombination and the transverse comoving distance to that redshift respectively. The correlation between these parameters are determined from a set of early Universe results based on the \texttt{CamSpec} likelihood \cite{Lemos:2023xhs}. These CMB priors compress the full likelihood into a high-redshift calibration for low-redshift probes like BAO and this compressed information is independent of late-time dark energy evolution. Numerical details of this implementation and the corresponding covariance matrix can be found in Appendix A of \cite{DESI:2024mwx}. We refer to this dataset as CMB.   
\end{itemize}

In our analysis, we use the following combination of datasets: 
\begin{itemize}
    \item[(a)] BAO + CC + CMB + PP, 
    \item[(b)] BAO + CC + CMB + DD.
\end{itemize} 
The constraints on the model parameters are obtained by maximizing the total log-likelihood given by
\begin{align}
    -2 \log \mathcal{L_{\rm tot}} = \chi^2_{\rm tot},
\end{align}
where $\chi^2_{\rm tot}$ is the total chi-squared i.e 
\begin{align}
    \chi^2_{\rm tot} &= \chi^2_{\rm CC} + \chi^2_{\rm BAO} + \chi^2_{\rm CMB} + \chi^2_{\rm PP} \quad \text{ for (a)}, \nonumber \\
    \chi^2_{\rm tot} &= \chi^2_{\rm CC} + \chi^2_{\rm BAO} + \chi^2_{\rm CMB} + \chi^2_{\rm DD} \quad \text{ for (b)}. \nonumber
\end{align} 
We compute 1D and 2D posterior probability distribution from the MCMC samples along with the median and $1\sigma$ estimates. The performance of the models with respect to the flat $w_0w_a$CDM is analyzed by calculating the Akaike Information Criterion (AIC) and the Bayesian Information Criterion (BIC) defined as \cite{Akaike:1974vps,Schwarz:1978tpv,Trotta:2008qt}
\begin{align}
    AIC &= -2 \log \mathcal{L_{\rm tot}} + 2k, \\
    BIC &= -2 \log \mathcal{L_{\rm tot}} + k\log N,
\end{align}
where $k$ is the number of effective free parameters and $N$ is the number of datapoints. Model preference can be assessed by calculating the difference in information criteria (IC) between the considered models
\begin{align}
    \Delta IC = IC_{i} - IC_{min}.
\end{align}
In general, a model with lower value of $\Delta IC $ s preferred. A model with $\Delta IC < 2$ is said to have strong support, $2<\Delta IC < 7$ shows a weak support and models with $\Delta IC > 10$ are disfavored.

\begin{table*}[]
\begin{tabular}{wl{4cm}wc{2.5cm}wc{2.5cm}wc{2.5cm}wc{2.5cm}wc{2.5cm}}
\hline
Model/Dataset & $\chi^2$ & AIC & BIC & $\Delta$AIC & $\Delta$BIC \\ \hline
\hline
\rule{0pt}{3ex}
\textbf{CC + BAO + CMB + DD } &  &  &  &  &  \\
\rule{0pt}{3ex}
$w_0w_a$CDM & 1656.59 & 1666.59 & 1694.21 & $-$ & 4.27 \\
\rule{0pt}{3ex}
$\lambda_1$CDM & 1659.33 & 1667.33 & 1689.94 & 0.74 & $-$ \\
\rule{0pt}{3ex}
$\lambda_1\lambda_2$CDM & 1657.65 & 1667.65 & 1695.26 & 1.06 & 5.32 \\
\\[0.1mm]
\hline
\rule{0pt}{3ex}
\textbf{CC + BAO + CMB + PP } &  &  &  &  &  \\
\rule{0pt}{3ex}
$w_0w_a$CDM & 1430.77 & 1440.77 & 1467.73 & $-$ & 5.37 \\
\rule{0pt}{3ex}
$\lambda_1$CDM & 1432.80 & 1440.80 & 1462.36 & 0.03 & $-$ \\
\rule{0pt}{3ex}
$\lambda_1\lambda_2$CDM & 1432.70 & 1442.70 & 1469.65 & 1.93 & 7.29 \\
\\[0.1mm]
\hline
\end{tabular}
\caption{Model selection criteria and relative differences with respect to the baseline for the considered models. We see that for AIC the $w_0w_a$CDM model shows a slight preference over the $\lambda_1$CDM model while for the BIC the $\lambda_1$CDM is strongly favoured. }
\label{tab:modcomp}
\end{table*}

\begin{figure*}[]
   \begin{center}
        \includegraphics[width=8.5cm]{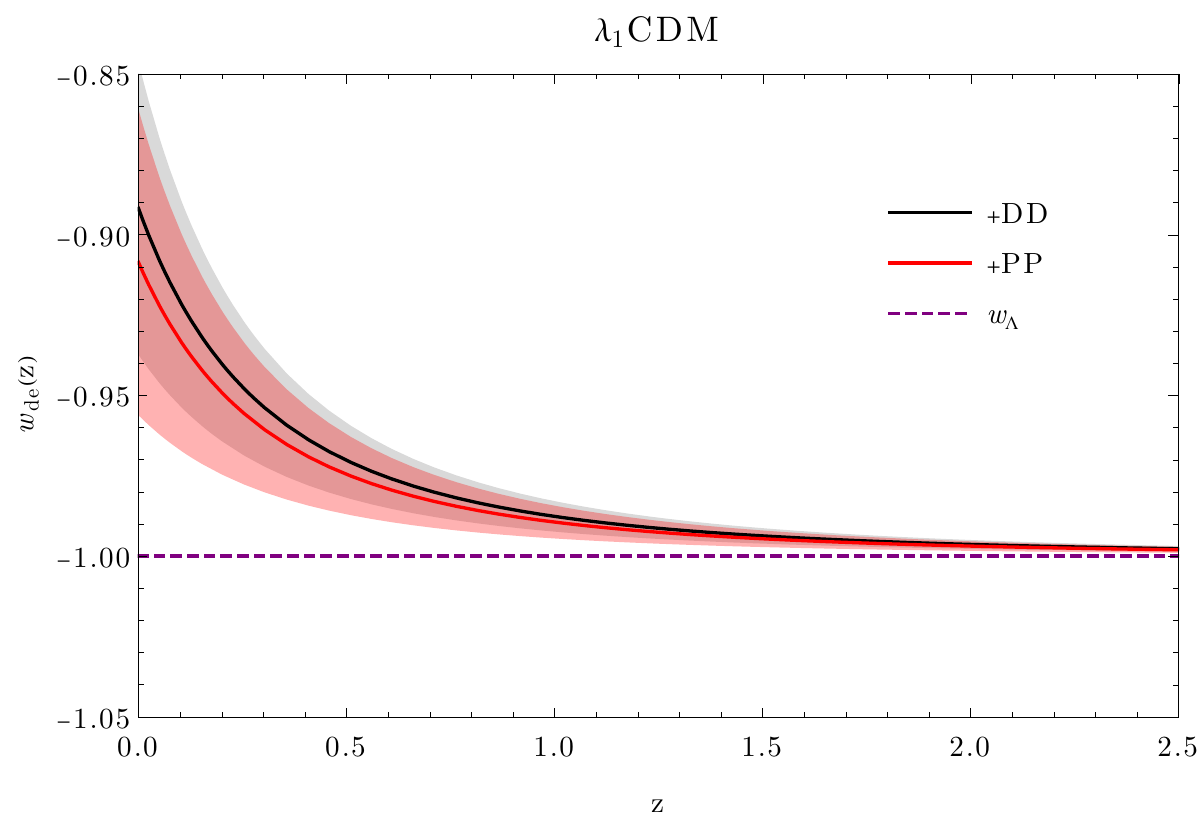}
        \includegraphics[width=8.5cm]{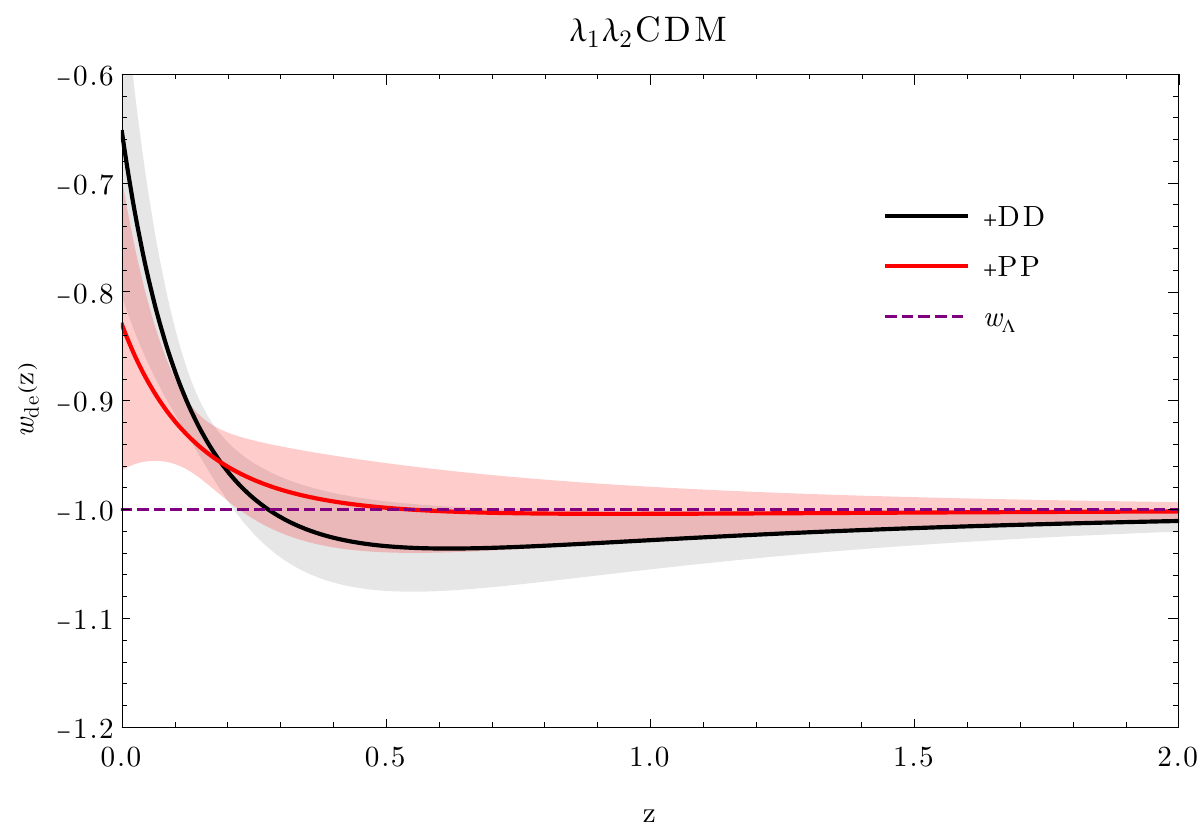}
  \end{center}
    \caption{Reconstructed equation of state $w_{de}$ for the $\lambda_1$CDM (left) and the $\lambda_1\lambda_2$CDM (right) models as a function of the redshift $z$. Here the black and red thick lines correspond to the best-fit values of the model parameters and the shaded region is 1$\sigma$ region for the respective curves.}\label{figeos}
\end{figure*}

\subsection{Results}

In this section, we discuss the results the parameter inference procedure. In Table \ref{tab:med1sig}, we present the median and $1\sigma$ uncertainty bounds of inferred parameter values of $w_0w_a$CDM, $\lambda_1$CDM and $\lambda_1\lambda_2$CDM models for the two dataset combinations outlined above. In Fig. \ref{figcont}, we show the one-dimensional posterior probability distributions and two-dimensional $68\%$ and $95\%$ confidence level contours for the free parameters of the model.  

For both dataset combinations, the inferred value of the Hubble constant remains remarkably stable across all models, with $H_0 \simeq 67.3 - 67.7  \ \text{km.s}^{-1}\text{Mpc}^{-1}$, and consistent at the sub-percent level. Similarly, the matter density parameter is tightly constrained to $\Omega_m \simeq 0.306 - 0.311$, with negligible dependence on the underlying dark energy parametrization.

The $\lambda_1$CDM model shows a pronounced deviation from the $\Lambda$CDM limit with $|\lambda_1| \lesssim 0.03$. The parameter $\lambda_1$ is constrained to be negative at the level of $\sim 2 - 3\sigma$ across both dataset combinations, indicating a mild but non-negligible tension with $\Lambda$CDM but no phantom behavior.

For the $\lambda_1\lambda_2$CDM model, both parameters remain consistent with zero within their respective uncertainties, and no statistically significant detection of additional dynamical degrees of freedom is observed. The two-parameter extension shows a clear degeneracy between $\lambda_1$ and $\lambda_2$, as evidenced by the elongated and tilted confidence contours in the right panel of Fig. \ref{figcont}. This indicates that current data primarily constrain a combination of these parameters rather than each independently. Addition of the higher order correction term does not lead to a substantial shift in the cosmological parameters or a tightening of constraints. The results obtained using the +DD and +PP supernova compilations are mutually consistent, reinforcing the robustness of these conclusions.

The model comparison based on the information criteria described in the previous section is summarized in Tab.~\ref{tab:modcomp}, where the standard $w_0w_a$CDM model is taken as a reference model. For both dataset combinations, the differences in the AIC remain small, with $\Delta AIC\lesssim2$ for all models. The $\lambda_1$CDM model yields $\Delta AIC\sim0.03 - 0.74$ across dataset combinations, indicating that it performs comparably to the $w_0w_a$CDM model despite having a smaller parameter space. On the other hand, the $\lambda_1\lambda_2$CDM model, which introduces an additional parameter, shows slightly larger values of $\Delta AIC$, still remaining within the range of statistically indistinguishability from the reference model.

We see a more pronounced distinction emerging when we consider the BIC. Due to its stronger penalization of model complexity, the BIC consistently favors the simpler $\lambda_1$CDM model over both the $\lambda_1\lambda_2$CDM model and the reference $w_0w_a$CDM model. In both dataset combinations, $\lambda_1$CDM yields the lowest BIC, while the two-parameter model is disfavored with $\Delta BIC \gtrsim 5$. However, this information is not reliable as the two deviation parameters are highly degenerate.

Finally in Fig.~\ref{figeos} we show the reconstructed equation of state parameter, $w_{de}$ of the dynamical dark energy component for both the models where the shaded regions show $1\sigma$ uncertainty bounds. As expected, the dynamical nature of the dark energy component is evident in these plots. By construction, the equation of state of $\lambda_1$CDM cannot cross the phantom-divide and it approaches $-1$ as $z$ increases. On the other hand, we see a phantom crossing behavior for the $\lambda_1\lambda_2$CDM model as the best-fit values of $\lambda_1$ and $\lambda_2$ have opposite signs. However, in view of the previous analysis, it is worth questioning whether such a phantom behavior is necessary with the currently available data, since the single parameter model $\lambda_1$CDM appears to provide a better fit.

\section{Discussions}\label{sec5}

We constructed a phenomenological framework to explain dynamical dark energy as a consequence of departures from GR in the low density regime from the Markov-Mukhanov action. We showed that the cosmological constant $\Lambda$ is naturally obtained as the first order term appearing in the expansion of the matter-gravity coupling, while considering higher order terms naturally leads to a dynamical dark energy component. We showed how the phantom crossing tentatively observed by DESI can be naturally explained by considering the two next leading order terms in the expansion.

It is worth mentioning that an early phantom behavior, that recently crossed towards quintessence-like, which aligns with the application of the $w_0w_a$CDM parametrization to DESI's observations \cite{DESI:2024mwx,DESI:2025fii}, can be recovered only for the $\lambda_1\lambda_2$CDM as shown by both fits obtained in Table \ref{tab:med1sig}.

The CPL parametrization assumes an expansion for the dark energy equation of state parameter close to today of the form $w_{de}=w_0+w_a(1-a)$, while our approach introduces the expansion in the energy density at the level of the action. Here we wish to emphasize that Markov-Mukhanov models discussed above and the CPL can not be directly mapped onto each other. First of all, it is immediately clear that one can not map the $\lambda_1$CDM model to the CPL formalism because the former has only one additional free parameter while the latter has two. At the same time, while it is always possible to write the $\lambda_1\lambda_2$CDM model close to $a=1$ in the form of the CPL equation of state, the values of $w_0$ and $w_a$ obtained from the best fit for the $\lambda_1\lambda_2$CDM model do not correspond to the values obtained by fitting the CPL equation of state directly. This is especially important since as a consequence the two approaches provide different times for the occurrence of the phantom crossing (which is obviously model dependent).

More interestingly, the $\lambda_1$CDM model, despite having one fewer degree of freedom than the CPL parametrization, performs comparable to CPL while potentially avoiding the phantom crossing by construction.  
The $\lambda_1$CDM effectively retains two parameters governing the evolution of the dark-energy equation of state: its present-day value $w_{de}(z=0)$ and a next-order contribution that characterizes its redshift evolution, both of which depend on $\lambda_1$. This model is restricted to either phantom or quintessence (without any crossing) at all times. This suggests the possibility that the preference for phantom behavior reported in the CPL analysis and some other non-parametric reconstructions from the DESI data such as in \cite{DESI:2025fii}, may not necessarily show robust evidence for phantom crossing.

More precise data is needed in order to determine whether the currently observed phantom crossing is due to the nature of the parametrization or the existence of additional terms.
This is also apparent from the degeneracy of the fit for the $\lambda_1\lambda_2$CDM model which at present does not rule out the absence of phantom crossing.

\section*{Acknowledgment}
This research was supported by Nazarbayev University Faculty Development Competitive Research Grant Program No. 040225FD4737 `Modifications of General Relativity in the strong curvature regime and their implications for black holes and cosmology'.

\bibliography{ref}

\end{document}